\documentclass[12pt]{article}
\usepackage{fullpage,amssymb,amsmath}
\usepackage[dvips]{epsfig}

\def\##1{{\bf #1}}
\def\=#1{\underline{\underline{#1}}}

\def\+#1{\underline{\bf #1}}
\def\*#1{\underline{\underline{\bf #1}}}

\def\r#1{(\ref{#1})}
\def\l#1{\label{#1}}
\def\c#1{\cite{#1}}

\def\le{\left(}
\def\ri{\right)}
\def\les{\left[}
\def\ris{\right]}

\def\.{\mbox{ \tiny{$^\bullet$} }}

\def\eps{\epsilon}

\def\hcm{\mbox{\tiny{HCM}}}

\def\Br{\mbox{\tiny{Br}}}
\def\MG{\mbox{\tiny{MG}}}

\def\eBr{\eps^{\Br}_{\hcm}}
\def\eMGa{\eps^{\MG }_{\hcm/a}}
\def\eMGb{\eps^{\MG }_{\hcm/b}}

\begin{document}

\LARGE
\begin{center}
{\bf A limitation of the Bruggeman  formalism for homogenization}

\vspace{10mm} \large

Tom G. Mackay\footnote{Corresponding Author. Fax: + 44 131
650 6553; e--mail: T.Mackay@ed.ac.uk.}\\
{\em School of Mathematics,
University of Edinburgh, Edinburgh EH9 3JZ, UK}\\
\vspace{3mm}
 Akhlesh  Lakhtakia\footnote{Fax:+1 814 865 99974; e--mail: akhlesh@psu.edu}\\
 {\em CATMAS~---~Computational \& Theoretical
Materials Sciences Group\\ Department of Engineering Science and
Mechanics\\ Pennsylvania State University, University Park, PA
16802--6812, USA}

\end{center}

\vspace{4mm}

\normalsize

\begin{abstract}

The Bruggeman formalism provides an estimate of the effective
permittivity of a particulate composite medium comprising two component mediums.
The Bruggeman estimate is required to lie within the Wiener bounds and
the Hashin--Shtrikman bounds. Considering
the homogenization of
weakly dissipative component mediums 
 characterized by relative
permittivities with  real parts of opposite signs,
we show that the
Bruggeman estimate may not be not  physically reasonable when the component mediums
are weakly dissipative; furthermore,
both the Wiener bounds and the Hashin--Shtrikman bounds  exhibit strong
resonances.

\end{abstract}


\section{Introduction}

Metamaterials in the form of  particulate composite mediums are currently of
considerable scientific and technological interest \c{Walser}.
Provided that wavelengths are sufficiently long compared with the
length scales of inhomogeneities,  such a metamaterial may be
envisaged as a homogenized composite medium (HCM), arising from
two  homogeneous component mediums \c{L96, Mackay03}. HCMs with especially
interesting properties may be conceptualized if the real parts of
the
 relative permittivities (and/or relative permeabilities)  of the two
 component mediums have opposite signs \c{Lijimmw}.
This possibility arises for metal--in--insulator 
dielectric composites \c{Sherwin, MLW01} and 
has recently become  feasible with the fabrication
of dielectric--magnetic materials displaying a negative index of
refraction in the microwave frequency range \c{Shelby,Smith}.

Over many years,  several  theoretical formalisms have been developed in order
 to estimate the effective constitutive parameters of particulate composite mediums \c{L96}.
 In particular, the  Maxwell Garnett and the Bruggeman homogenization formalisms
 have been  widely  used \c{Ward}. Generally, the Maxwell Garnett formalism
 is seen to hold  only for  dilute composite mediums \c{MG}.\footnote{The restriction on the applicability of
 the Maxwell Garnett formalism to
 dilute composite mediums 
 generally emerges from comparison with experimental data \c{Ward}. As the particulate volume fraction increases,
 the distribution of particles begins to  lose the randomness which is inherent to the theory
 of the Maxwell Garnett formalism \c{Lijaem}. However, the restriction
  could be bypassed if the distribution of particles in a composite medium continues to lack order even under densification,
 which thought underlies the {\em random unit cell\/} approach developed by Smith
 and colleagues \c{Sm1, Niklasson,Sm2}. An anonymous reviewer has suggested that self--assembly
 techniques \c{Zhang} can yield randomness even at large particulate volume fractions, and could therefore extend
 the applicability of the Maxwell Garnett formalism.}
 More
 widely applicable is the Bruggeman formalism that  was initially
 founded on the intuition that the total
 polarization field is zero throughout the HCM \c{Brugge}.
A rigorous basis for the Bruggeman formalism is  also available,
within the framework of the strong--permittivity--fluctuation theory (SPFT)
\c{TK81, MLW00}.

Estimates of HCM constitutive parameters generated by
homogenization formalisms may
be required to lie within certain bounds. In particular, the
Wiener bounds \c{Wiener, Aspnes} and the Hashin--Shtrikman bounds \c{HS} are often invoked.
 The Hashin--Shtrikman bounds coincide with the constitutive
 parameter estimates of the Maxwell Garnett homogenization
 formalism \c{Aspnes}. The applicability of  theoretical bounds on the HCM
permittivity has recently been the focus of attention  for composites
 specified by  relative permittivities with
positive--valued
real parts \c{Sihvola}.

In this communication, we consider the application of the
Bruggeman formalism, together with the Wiener and
Hashin--Shtrikman bounds, to isotropic dielectric HCMs which arise
from  component mediums characterized by 
 complex--valued relative
permittivities whose real parts have opposite signs.
This is scenario is typical of metal--in--insulator HCMs \c{Aspnes, Milton}, for example.
By duality,  our analysis  extends to 
 isotropic magnetic HCMs. It also extends to 
isotropic dielectric--magnetic HCMs, because the permeability and the permittivity are
then independent of each other in the   Bruggeman formalism \c{Kampia} (as also in the
Maxwell Garnett formalism \c{Lak92}). Therefore, our findings
 are very relevant to the application of 
homogenization formalisms \c{Lijimmw} to mediums
displaying negative index of refraction \c{LMW03}, for example.
Furthermore, the implications of our mathematical study extend beyond 
the Bruggeman formalism to the SPFT
as well \c{Mackay03}.

A note on notation: An $\exp(-i\omega t)$ time--dependence is
implicit in the following sections; and
the real and imaginary parts of
complex--valued quantities are denoted by
 $\mbox{Re} \le \. \ri$ and  $\mbox{Im} \le \. \ri$, respectively.

\section{Theory}

\subsection{Bruggeman equation}

Consider the homogenization of two isotropic dielectric component
mediums labelled $a$ and   $b$.  Let their
relative permittivities
be denoted by $\eps_a$ and $\eps_b$, respectively. For later
convenience, we define
\begin{equation}
 \delta = \left\{
 \begin{array}{ccc}
 \displaystyle
 \frac{\eps_a}{\eps_b} & \quad \mbox{if} \quad  & \eps_a, \eps_b \in
 \mathbb{R},\\ && \\
 \displaystyle
\frac{\mbox{Re} \le \eps_a \ri }{\mbox{Re} \le \eps_b \ri} & \quad
\mbox{if} \quad  & \eps_a, \eps_b \in
 \mathbb{C}.
\end{array}
\right.
\end{equation}

The Bruggeman estimate of the HCM relative permittivity, namely
$\eBr$, is provided implicitly via the relation \c{Ward}
\begin{eqnarray}
&& \eBr = \frac{f_a \eps_a \le \eps_b + 2 \eBr \ri + f_b \eps_b
\le \eps_a + 2 \eBr \ri }{f_a  \le \eps_b + 2 \eBr \ri + f_b \le
\eps_a + 2 \eBr \ri}, \l{Br}
\end{eqnarray}
wherein  $f_a$ and $f_b$ are the  respective volume fractions of
component mediums $a$ and $b$, and the particles of both component
mediums are assumed to be spherical. The Bruggeman  equation \r{Br}
emerges naturally within the  SPFT framework \c{Mackay03}. A rearrangement of
\r{Br} gives the quadratic equation
\begin{equation}
2 \le \eBr \ri^2 + \eBr \les \eps_a \le f_b - 2f_a \ri + \eps_b
\le f_a - 2f_b \ri \ris - \eps_a \eps_b = 0. \l{quadratic}
\end{equation}
Only those $\eBr$--solutions of \r{quadratic} are valid under the principle of causality as encapsulated
by the Kramers--Kronig relations \c{BH} which conform to the restriction  $\mbox{Im}\,\le
\eBr \ri \ge 0$.

Let $\Delta$ be the  discriminant of the quadratic equation \r{quadratic}; i.e.,
\begin{equation}
\Delta = \les \eps_a \le f_b - 2f_a \ri + \eps_b \le f_a - 2f_b
\ri \ris^2 + 8 \eps_a \eps_b . \l{Delta1}
\end{equation}
Since $f_b = 1 - f_a$, we may express \r{Delta1} as
\begin{equation}
\Delta =  9 f^2_a \le \eps_a - \eps_b \ri^2 -6 f_a \le \eps^2_a -3
\eps_a \eps_b + 2 \eps^2_b \ri + \le \eps_a + 2 \eps_b \ri^2.
\end{equation}
An insight into the applicability of the Bruggeman formalism may
be gained by considering the $f_a$--roots of the equation $\Delta =
0$; these are as follows:
\begin{equation}
\left.
f_a \right|_{\Delta = 0} =  \frac{\eps^2_a -3 \eps_a
\eps_b + 2 \eps^2_b \pm 2 \sqrt{2} \sqrt{ - \eps_a \eps_b \le
\eps_a - \eps_b \ri^2}}{3  \le \eps_a - \eps_b \ri^2}.
\end{equation}
On restricting attention  to nondissipative component
mediums (i.e., $\eps_{a,b} \in \mathbb{R}$),  it is clear that  $ \left. f_a \right|_{\Delta=0}$ are
complex--valued if $\delta
> 0$. Consequently,  $\Delta > 0$ which implies that
$\eBr \in \mathbb{R}$. On the other hand,
  $ \left. f_a \right|_{\Delta=0}$
 are real--valued if $\delta < 0$.   Thus, the Bruggeman estimate
$\eBr$ for $\delta < 0 $ may be  complex--valued with nonzero imaginary part,
even though neither component medium is dissipative.

\subsection{Bounds on the HCM relative permittivity}

Various bounds on the  HCM relative permittivity have been
developed. Two of the most widely used are the Wiener bounds
\c{Wiener, Aspnes}
\begin{equation}
\left.
\begin{array}{l}
W_\alpha = \le \frac{f_a}{\eps_a} + \frac{f_b}{\eps_b} \ri^{-1} \\
\vspace{-4mm} \\
W_\beta = f_a \eps_a + f_b \eps_b
\end{array}
\right\}\,\l{Wiener}
\end{equation}
 and the Hashin--Shtrikman
bounds \c{HS}
\begin{equation}
\left.
\begin{array}{l}
{HS}_\alpha = \eps_b + \frac{3 f_a \eps_b \le \eps_a - \eps_b
\ri}
{\eps_a + 2 \eps_b - f_a \le \eps_a - \eps_b \ri} \\ \vspace{-2mm} \\
{HS}_\beta = \eps_a + \frac{3 f_b \eps_a \le \eps_b - \eps_a
\ri} {\eps_b + 2 \eps_a - f_b \le \eps_b - \eps_a \ri}
\end{array}
\right\}\,. \l{Hashin}
\end{equation}
While both the Wiener bounds and the Hashin--Shtrikman bounds were
originally derived for real--valued constitutive parameters,
generalizations to complex--valued constitutive parameters have
been established \c{Milton}.

The Hashin--Shtrikman bound $\mbox{HS}_\alpha$ is
equivalent to the Maxwell Garnett  estimate of the HCM relative
permittivity $\eMGa$  based on spherical particles of component
medium $a$ embedded in the host component medium $b$.  Similarly,
$\mbox{HS}_\beta$ is equivalent to the Maxwell Garnett  estimate
of the HCM relative permittivity $\eMGb$ based on spherical
particles of component medium $b$ embedded in the host component
medium $a$. The estimate $\eMGa$ is valid for $f_a  \lesssim 0.3$,
whereas the estimate $\eMGb$ is valid for $f_b \lesssim 0.3$; but see the
footnote in Section 1.

To gain insights into the asymptotic behaviour of the Wiener and
Hashin--Shtrikman bounds, let us again restrict attention to the
case of nondissipative component mediums (i.e., $\eps_{a,b} \in
\mathbb{R}$). From \r{Wiener}, we see that $W_\beta$ remains
finite for all values of $\delta$, but $W_\alpha$ may become
infinite for $\delta < 0$ since
\begin{equation}
| W_\alpha | \rightarrow \infty \qquad \mbox{as} \qquad \delta
\rightarrow  - \frac{f_a}{f_b}\, . \l{Wa_lim}
\end{equation}
In a similar vein, from \r{Hashin} we find that
\begin{equation}
| HS_\alpha | \rightarrow \infty \qquad \mbox{as} \qquad \delta
\rightarrow \frac{f_b - 3}{f_b}\,; \l{HSa_lim}
\end{equation}
thus, for all values of  $\delta < -2  $ there exists a value of
$f_b \in (0,1)$ at which $HS_\alpha$ is unbounded. Analogously,
\begin{equation}
| HS_\beta | \rightarrow \infty \qquad \mbox{as} \qquad \delta
\rightarrow \frac{f_a}{f_a - 3}; \l{HSb_lim}
\end{equation}
so  we can always
find a value of $f_a \in (0,1)$ at which $HS_\beta$ is unbounded, provided that  $\delta \in ( -\frac{1}{2}, 0 ) $.

\section{Numerical results}

Let us now present,  calculated values of the HCM relative
permittivity $\eBr$,  along with the corresponding values of the
bounds $W_{\alpha, \beta}$ and $HS_{\alpha, \beta}$,     for some representative examples. Both nondissipative
and dissipative HCMs are considered for $\delta = \pm 3$.

\subsection{Nondissipative component mediums}

 The effects
of dissipation may be  very clearly appreciated through first 
considering the idealized situation  wherein the components mediums
are nondissipative \c{vandeHulst}. Furthermore, although the absence of dissipation is unphysical due to
the dictates of causality \c{BH},
weak dissipation in a particular spectral regime 
is definitely possible and is then often ignored \cite[Sec.2.5]{Ward}.

 Thus, it is instructive to begin with  the commonplace
 scenario wherein both $\eps_a > 0$ and $\eps_b > 0$.
For example, let
 $\eps_a = 6$ and $\eps_b = 2$.
In Figure~1, $\eBr$ is plotted against $f_a$, along with the corresponding
Wiener bounds
  $W_{\alpha, \beta}$
and  Hashin--Shtrikman bounds  $HS_{\alpha, \beta}$. The latter bounds are stricter than the former
bounds in the sense that
\begin{equation}
W_\alpha < HS_\alpha <  \eBr < HS_\beta < W_\beta. \l{inequ}
\end{equation}
The close agreement between $\eBr$ and the lower Hashin--Shtrikman
bound $HS_\alpha$ at low volume fractions $f_a$ is indicative of
the fact that $HS_\alpha \equiv \eMGa$.
 Similarly,
 $\eBr$ agrees closely with the upper
Hashin--Shtrikman bound $HS_\beta$ at high  values of $f_a$ since
$ HS_\beta \equiv \eMGb$.

A markedly  different situation develops if the  real--valued $\eps_a $ and $\eps_b $ have opposite signs.
For example, the  values  of $\eBr$ calculated  for $\eps_a = -6$ and $\eps_b = 2$ are graphed against $f_a$  in
Figure~2, together with the corresponding Wiener and
Hashin--Shtrikman bounds. The Bruggeman estimate $\eBr$ is
complex--valued with nonzero imaginary part for $f_a \lesssim
0.82$. This estimate
 is not physically reasonable. The Bruggeman homogenization formalism~---~unlike the
SPFT which is its natural generalization~---~has no mechanism for
taking coherent scattering losses into account. Furthermore, no account has been taken
in the Bruggeman equation \r{Br} for the finite size of the particles \c{Lijaem,PLS,Shan96}. Therefore, the
Bruggeman estimate of the HCM relative permittivity is required to
be real--valued if the component mediums are nondissipative.

While $\eBr$ in Figure~2 is complex--valued,
 the Wiener bounds
and the Hashin--Shtrikman bounds are both real--valued. In
accordance with \r{Wa_lim}, we see that $| W_\alpha | \rightarrow
\infty$ as $f_a \rightarrow \frac{3}{4}$. Similarly, $|
HS_\alpha | \rightarrow \infty$ in the limit  $f_a \rightarrow
\frac{1}{4}$, as may be anticipated from \r{HSa_lim}. Furthermore, since 
$HS_\alpha \equiv \eMGa$, the Maxwell Garnett formalism is clearly
inappropriate here. We also observe that the inequalities \r{inequ}
which hold for $\delta > 0$, do not hold for $\delta < 0$.

\subsection{Weakly dissipative component mediums}

Let us now investigate  $\eBr$ and its associated bounds  when the
component mediums are dissipative; i.e.,  $\eps_{a,b} \in
\mathbb{C}$. We begin with those cases for which  $\delta > 0$:
 for
example,  we take $\eps_a = 6 + 0.3i$ and $\eps_b = 2 + 0.2i$. In
Figure~3, $\eBr$ is plotted against $f_a$, and the associated Wiener bounds
$W_{\alpha, \beta}$ and the Hashin--Shtrikman bounds $HS_{\alpha,
\beta}$ are also presented. The behaviour of the real parts of
$\eBr$, $W_{\alpha, \beta}$ and $HS_{\alpha, \beta}$
closely resembles that displayed in  the nondissipative example
of  Figure~1. In fact,
 the
following  generalization of \r{inequ} holds:
\begin{equation}
 \mbox{Re} \le W_\alpha  \ri < \mbox{Re} \le HS_\alpha
\ri < \mbox{Re} \le  \eBr \ri  < \mbox{Re} \le HS_\beta \ri  <
\mbox{Re}
\le W_\beta \ri .
 \l{inequ2}
\end{equation}
However, this ordering \r{inequ2}  does not extend to
the imaginary parts  of $\eBr$, $W_{\alpha, \beta}$ and
$HS_{\alpha, \beta}$.

Turning to the cases for $\delta < 0$, we let $\eps_a = -6 + 0.3i$ and
$\eps_b = 2 + 0.2i$, for example. The corresponding Bruggeman
estimate $\eBr$ is graphed as function of $f_a$, along with  the
Wiener bounds $W_{\alpha, \beta}$ and the Hashin--Shtrikman bounds
$HS_{\alpha, \beta}$ in Figure~4. Since $\mbox{Im} \le \eps_{a,b}
\ri \neq 0$, the real parts of $W_\alpha$ and $HS_\alpha$ remain
finite, unlike in the corresponding nondissipative scenario
presented in Figure~2.

However,  the real and imaginary parts of
$W_\alpha$ and $HS_\alpha$ exhibit strong resonances in the
vicinity of $f_a = \frac{3}{4}$ (for $W_\alpha$)  and $f_a =
\frac{1}{4}$ (for $HS_\alpha$).  These resonances become
considerably more pronounced if the degree of dissipation
exhibited by the component mediums is reduced. For example, in
Figure~5 the graphs corresponding to Figure~4 are reproduced for
 $\eps_a = -6 +
0.003i$ and $\eps_b = 2 + 0.002i$. We observe in particular that
$\mbox{Im} \le \eBr \ri > 1$ for $ 0.05 \lesssim f_a \lesssim 0.8$.
Thus, the Bruggeman estimate $\eBr$  vastly exceeds both the
Wiener bounds $W_{\alpha, \beta}$ and the Hashin--Shtrikman bounds
$HS_{\alpha, \beta}$ for a wide range of $f_a$.
 Since $\mbox{Im} \le \eps_{a,b} \ri
\leq 0.003$, the estimates of ${\rm Im}(\eBr)$ are clearly
unreasonable. 

Furthermore, since the real and imaginary parts of
$HS_{\alpha} \equiv \eMGa  $ exhibit  sharp resonances at $f_a = \frac{1}{4}$,
we may infer that the Maxwell Garnett formalism is inapplicable for $\delta < 0$.

\subsection{Highly dissipative component mediums}

On comparing Figures~4 and 5, we conclude that the Bruggeman
formalism, the Weiner bounds and the Hashin--Shtrikman bounds become increasing
inappropriate as the degree of dissipation decreases towards zero.  This means that
all three {\em could\/} be applicable rather well when the dissipation is not weak. 

Therefore, let us examine the scenario wherein the real and imaginary parts of the relative permittivities of the
component medium are of the same order of magnitude;
i.e., we take $\eps_a = -6 + 3i$ and $\eps_b = 2 + 2i$. The
corresponding plots of
the Bruggeman estimate $\eBr$ together with the 
 Wiener bounds $W_{\alpha, \beta}$ and the Hashin--Shtrikman bounds
$HS_{\alpha, \beta}$ are presented in  Figure~6.
The real and imaginary parts of the Bruggeman estimate are
physically plausible, and both lie within the Hashin--Shtrikman
bounds. The Hashin--Shtrikman bounds themselves do not exhibit
resonances, and the Weiner bounds do not exhibit strong resonances.
Accordingly, we conclude that many previously published results are
not erroneous, but caution is still advised.

\section{Discussion}

The Bruggeman homogenization formalism is well--established in the
context of isotropic dielectric HCMs, as well as more generally
\c{L96}. However, this formalism was shown in Section 3.2 to be
inapplicable  for HCMs which arise from two 
isotropic dielectric
component mediums, characterized by  relative permittivities $\eps_a$
and $\eps_b$, with
\begin{itemize}
 \item[(i)]
 $\mbox{Re} \le \eps_a \ri$ and $\mbox{Re} \le
\eps_b \ri$
having opposite signs;  and
\item[(ii)] $| \mbox{Re} \le \eps_{a,b} \ri |$ $ \gg $
$| \mbox{Im} \le \eps_{a,b} \ri |$.
\end{itemize}
Since the Bruggeman formalism  provides the comparison medium
which underpins the  SPFT, it may be inferred that the SPFT is
likewise not applicable  to the scenarios of (i) with (ii).

It is also demonstrated in Section 3.2 that both
 the Wiener bounds and the Hashin--Shtrikman bounds
can exhibit strong resonances when the component mediums are
characterized by (i) with (ii). In the vicinity of resonances,
these bounds clearly do not constitute tight bounds on the HCM
relative permittivity. As a direct consequence, the Maxwell
Garnett homogenization formalism, like the Bruggeman
homogenization formalism, is inapplicable to the scenarios of  (i)
with (ii). This limitation also extends to the recently developed
incremental \c{IMG} and differential \c{DMG} variants of the
Maxwell Garnett formalism.

If the component mediums are sufficiently dissipative then 
the Bruggeman formalism and the Hashin--Shtrikman bounds (and
therefore
also the Maxwell Garnett formalism) provide physically plausible 
estimates, despite the real parts of the component medium 
relative permittivities 
having opposite signs~---~as shown in Section 3.3. The explicit delineation
of the appropriate parameter range(s) for the Bruggeman formalism and
the Hashin--Shtrikman bounds is a matter for
 future investigation.
 
 Bounds can, of course, be violated by a formalism if the underlying conditions for
 the formalism are in conflict with those used for deriving the bounds. Sihvola \c{Sihvola}
 has catalogued the following conflicts: 
 \begin{itemize}
 \item[(a)] Bounds derived for nondissipative component mediums can be invalid for
 the real parts of either $\eMGa$ or $\eMGb$ for a composite medium containing dissipative component mediums.
 \item[(b)] Percolation cannot be cannot be captured by the Maxwell Garnett formalism \c{Sherwin,Sasta}. Hence, the Hashin--Shtrikman
 bounds, being based on the Maxwell Garnett formalism, can be violated by
 the Bruggeman estimate $\eBr$ for a percolative composite medium.
 \item[(c)] The derivations of bounds generally assume that the particles in a composite medium
 have simple shapes. If the particle shapes are complicated, the composite medium
 may display properties not characteristic of the either of the
 component mediums. For instance, magnetic properties can be displayed when the particles in a composite medium
 have complex shapes \cite{Lbel,PHRS}, even though the component mediums are nonmagnetic. Clearly,
 the magnetic analogs of  $W_\alpha$, $W_\beta$, ${HS}_\alpha$ and ${HS}_\beta$
 are then inapplicable.
 \item[(d)] $W_\alpha$, $W_\beta$, ${HS}_\alpha$, and ${HS}_\beta$ as well as their magnetic
 analogs are also invalid {\em prima facie\/} when the component
mediums exhibit  magnetoelectric properties
 \c{Mackay03,Lijaem,Mich00}. 
 \item[(e)] Bounds derived for electrically small particles become inapplicable with increasing frequency, due to
 the emergence 
 of finite--size effects \c{PLS}. Even the
 concept of homogenization becomes questionable with increasing electrical size \cite[p. xiii]{L96}.
 \end{itemize}
 In contrast, the bounds and the homogenization formalisms studied in this paper share the same premises; yet, a conflict
 arises in certain situations because the bounds exhibit resonance while the homogenization estimates do not.

\section{Concluding remarks}
As several conventional approaches to homogenization are not
appropriate to the HCMs arising from component mediums characterized
 by (i) with (ii),
 there is a  requirement for new theoretical
techniques to treat this case. 
This
requirement is all the more pressing, given the  growing scientific
and technological importance of new types of
 metamaterials
\c{Walser, LMW03}.

\bigskip\bigskip
\noindent {\bf Acknowledgement.} We thank two anonymous reviewers for comments that led
to the improvement of this paper.

\newpage

\begin{figure}[!ht]
\centering \psfull \epsfig{file=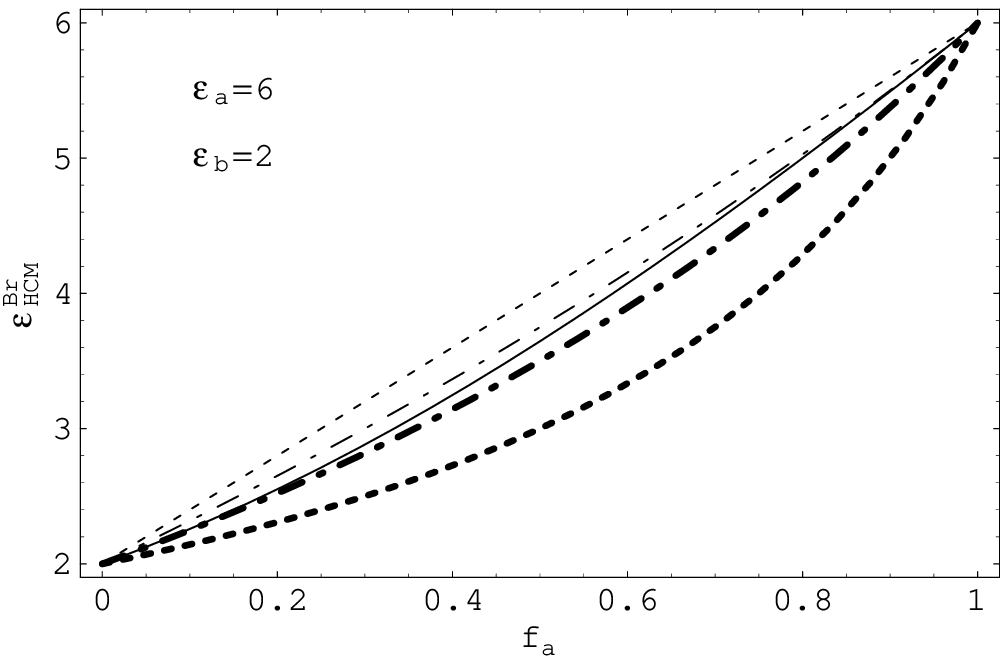,width=5.0in}
 \caption{\label{fig1}
The Bruggeman estimate  $\eBr$ (solid line) plotted against $f_a$
for $\eps_a = 6$ and $\eps_b = 2$. Also plotted are  the Wiener
bounds,   $W_\alpha$ (thick dashed line) and  $W_\beta$ (thin dashed line), and  the
Hashin--Shtrikman bounds, $HS_\alpha$ (thick broken dashed
line) and  $HS_\beta$ (thin broken dashed
line).
 }
\end{figure}

\newpage

\begin{figure}[!ht]
\centering \psfull \epsfig{file=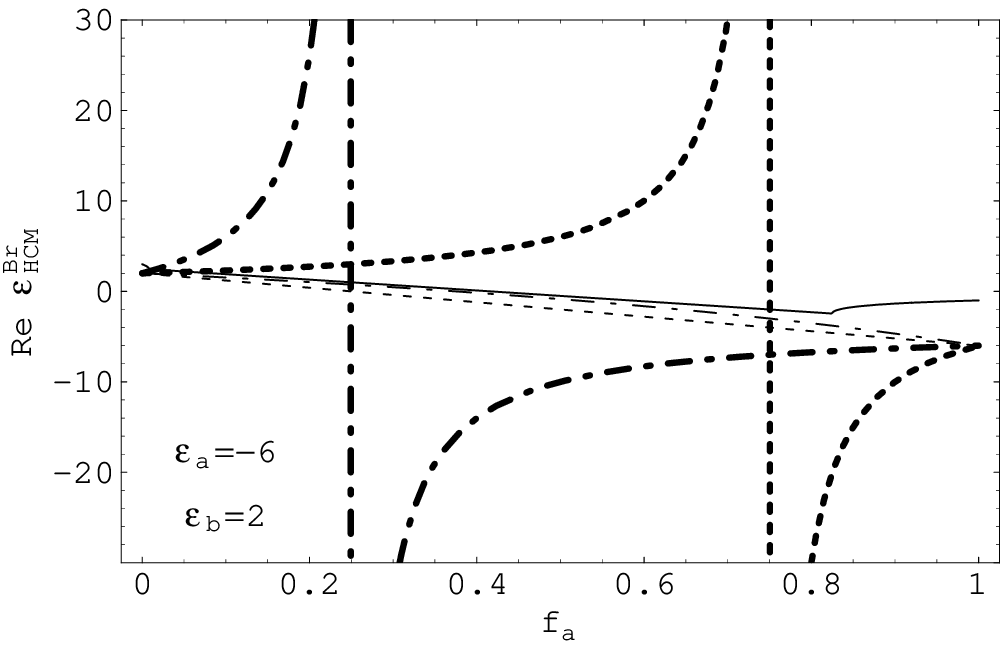,width=5.0in}
\epsfig{file=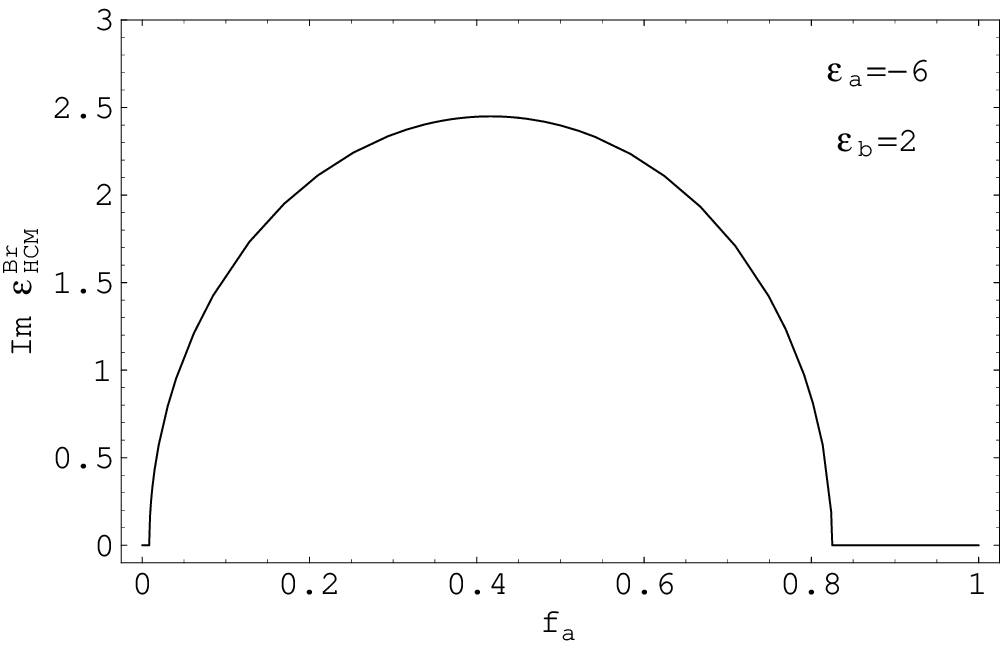,width=5.0in}
 \caption{\label{fig2}
The real (top) and imaginary (bottom) parts of the Bruggeman
estimate  $\eBr$ (solid line) plotted against $f_a$ for $\eps_a =
-6$ and $\eps_b = 2$. Also plotted are  the real parts of the
Wiener
bounds,   $W_\alpha$ (thick dashed line) and  $W_\beta$ (thin dashed line), and  the
Hashin--Shtrikman bounds, $HS_\alpha$ (thick broken dashed
line) and  $HS_\beta$ (thin broken dashed
line).
 The imaginary parts of  $W_{\alpha, \beta}$ and
$HS_{\alpha, \beta}$ are null--valued. }
\end{figure}

\newpage

\begin{figure}[!ht]
\centering \psfull \epsfig{file=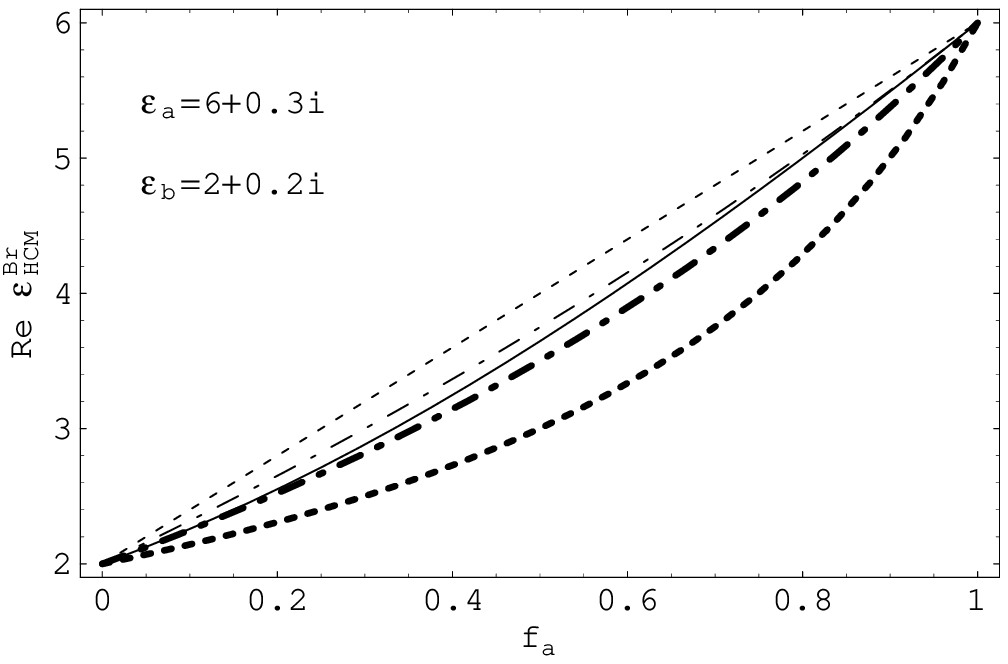,width=5.0in}
\epsfig{file=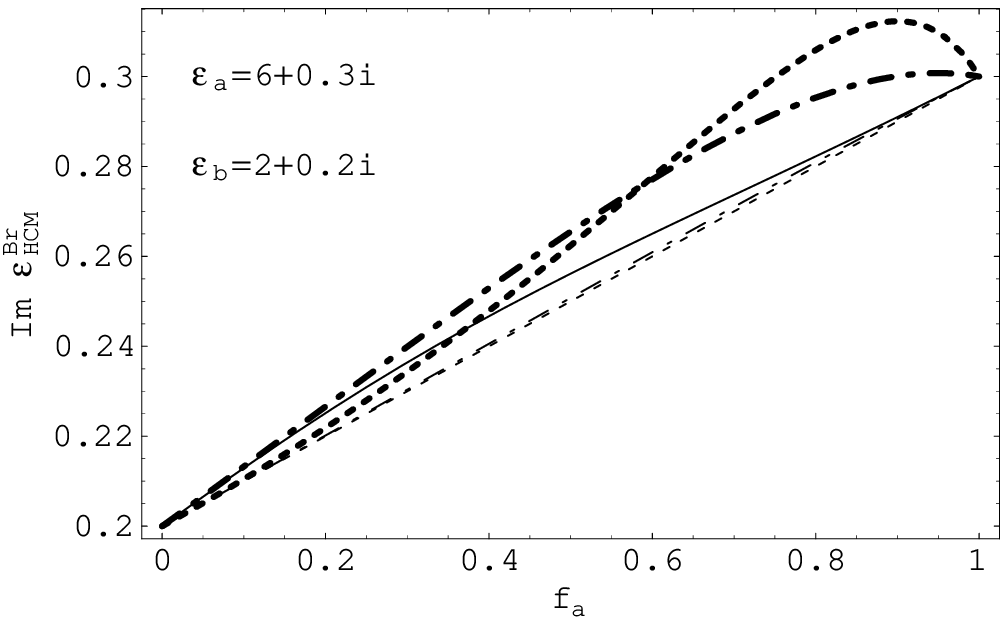,width=5.0in}
 \caption{\label{fig3}
 The  real (top) and imaginary (bottom) parts of the
 Bruggeman estimate  $\eBr$ (solid line) plotted against $f_a$
for $\eps_a = 6 + 0.3i$ and $\eps_b = 2 + 0.2i$.
 Also plotted are  the Wiener
bounds,   $W_\alpha$ (thick dashed line) and  $W_\beta$ (thin dashed line), and  the
Hashin--Shtrikman bounds, $HS_\alpha$ (thick broken dashed
line) and  $HS_\beta$ (thin broken dashed
line).
  }
\end{figure}

\newpage

\begin{figure}[!ht]
\centering \psfull \epsfig{file=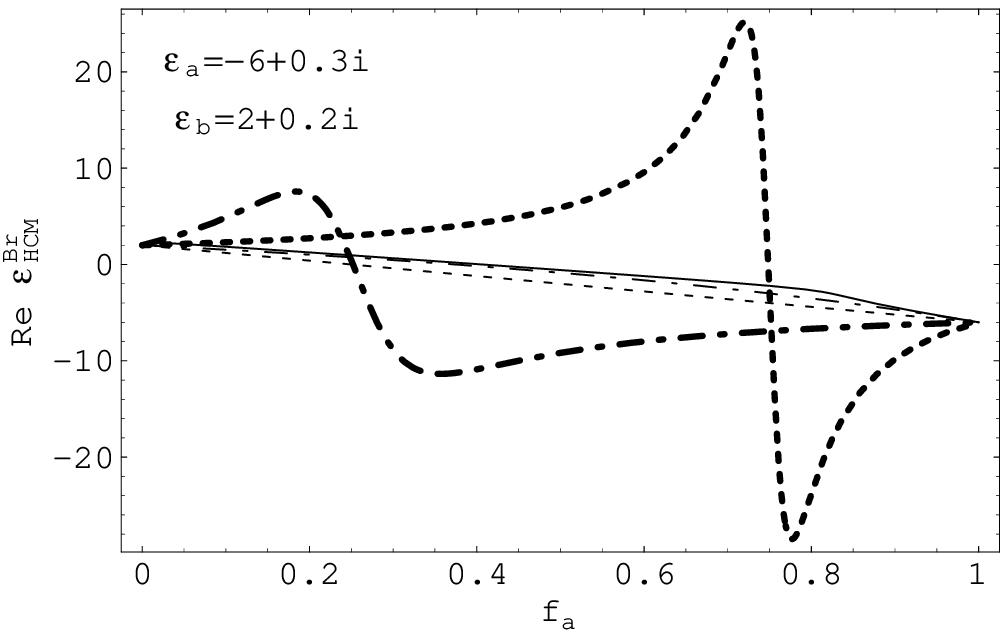,width=5.0in}
\epsfig{file=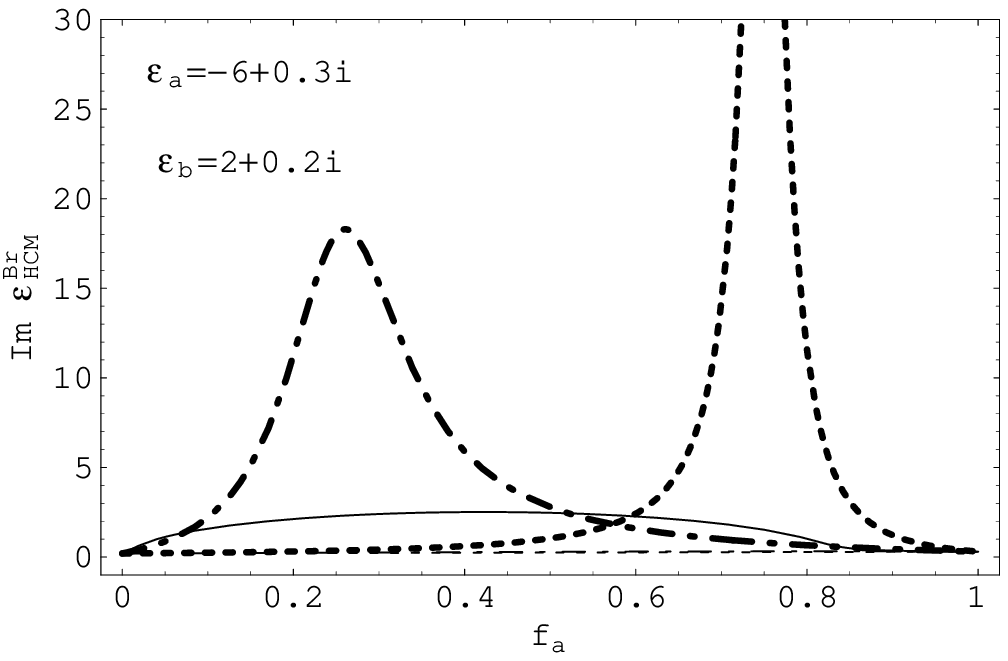,width=5.0in}
 \caption{\label{fig4}
As Figure~3 but for
 $\eps_a = -6 + 0.3i$ and $\eps_b = 2 + 0.2i$.
 }
\end{figure}

\newpage

\begin{figure}[!ht]
\centering \psfull \epsfig{file=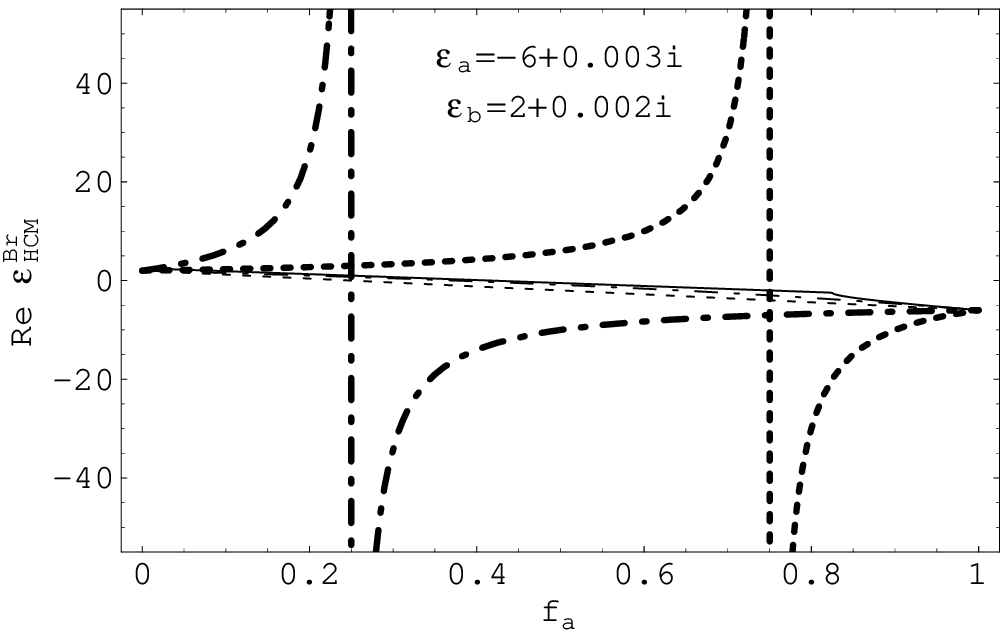,width=5.0in}
\epsfig{file=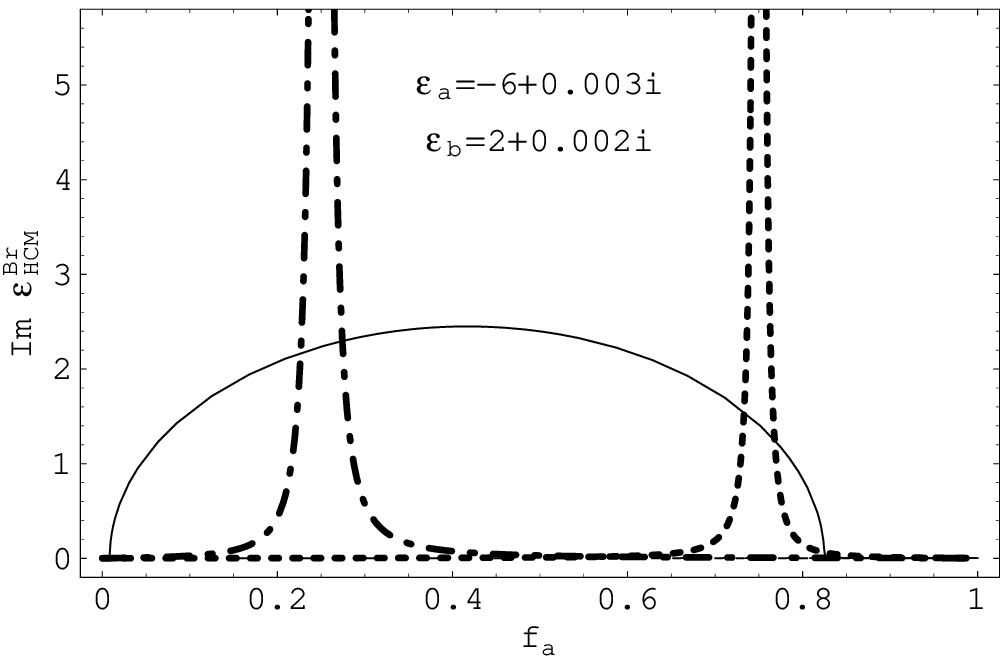,width=5.0in}
 \caption{\label{fig5}
As Figure~3 but for
 $\eps_a = -6 + 0.003i$ and $\eps_b = 2 + 0.002i$.
  }
\end{figure}
\newpage

\begin{figure}[!ht]
\centering \psfull \epsfig{file=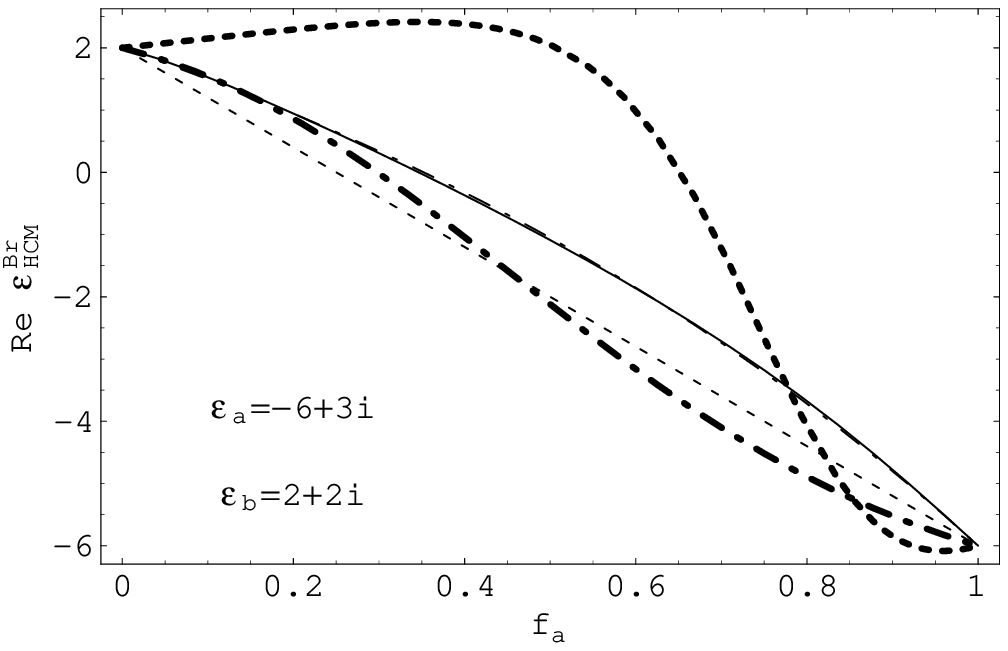,width=5.0in}
\epsfig{file=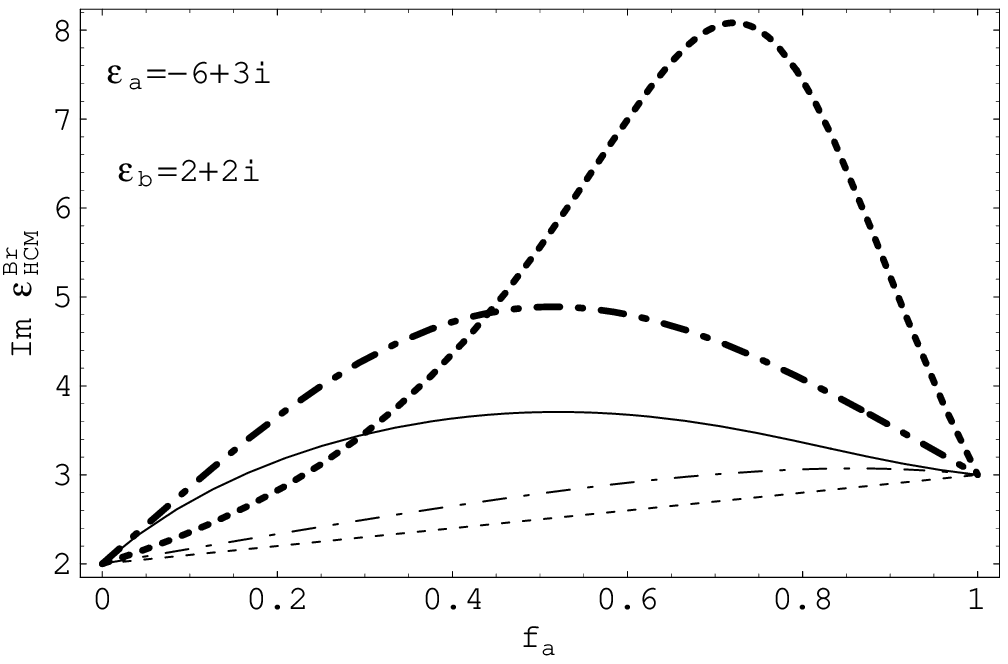,width=5.0in}
 \caption{\label{fig6}
As Figure~3 but for
 $\eps_a = -6 + 3i$ and $\eps_b = 2 + 2i$.
  }
\end{figure}

\end{document}